# Proxy Design: A Method for Involving Proxy Users to Speak on Behalf of Vulnerable or Unreachable Users in Co-Design

Anna Sigridur Islind [1*], Johan Lundin [2], Katerina Cerna [3],
Tomas Lindroth [4], Linda Åkeflo [5] & Gunnar Steineck [6]

*Corresponding author: islind@ru.is

[1] Department of Computer Science, Reykjavik University, Reykjavik, Iceland
[2] Department of Learning, Communication and IT, University of Gothenburg, Gothenburg, Sweden
[3] School of Information Technology, Halmstad University, Halmstad, Sweden
[4] Department of Informatics, University of Gothenburg, Gothenburg, Sweden
[5] Sahlgrenska University Hospital, Gothenburg, Sweden
[6] Division of Clinical Cancer Epidemiology, Sahlgrenska Academy, Gothenburg, Sweden

## Abstract

**Purpose:** Designing digital artifacts is not a linear, straightforward process. This is particularly true when applying a user-centered design approach, or co-design, with users who are unable to participate in the design process. Although the reduced participation of a particular user group may harm the end result, the literature on solving this issue is sparse. In this article, proxy design is outlined as a method for involving a user group as proxy users to speak on behalf of a group that is difficult to reach. The article investigates the following research question: *How can roleplaying be embedded in co-design to engage users as proxies on behalf of those who are unable to represent themselves?*
**Method:** The article presents a design ethnography spanning three years at a cancer rehabilitation clinic, where digital artifacts were designed to be used collaboratively by nurses and patients. The empirical data were analyzed using content analysis and consisted of 20 observation days at the clinic, six proxy design workshops, 21 telephone consultations between patients and nurses, and log data from the digital artifact.
**Findings:** The article shows that simulated consultations, with nurses roleplaying as proxies for patients ignited and initiated the design process and enabled an efficient in-depth understanding of patients. Moreover, the article reveals how proxy design as a method further expanded the design. The study findings illustrate: (1) proxy design as a method for initiating design, (2) proxy design as an embedded element in co-design and (3) six design guidelines that should be considered when engaging in proxy design.
**Originality:** The main contribution is the conceptualization of proxy design as a method that can ignite and initiate the co-design process when important users are unreachable, vulnerable or unable to represent themselves in the co-design process. More specifically, based on the empirical findings from a design ethnography that involved nurses as proxy users speaking on behalf of patients, the article shows that roleplaying in proxy design is a fitting way of initiating the design process, outlining proxy design as an embedded element of co-design.

# 1. Introduction

User involvement in the design process is widely recognized as an essential component in successfully designing and developing useful, usable, and user-friendly digital infrastructures. However, achieving user involvement is often challenging, particularly if a user group is unable to participate. In most cases, in-depth user involvement is facilitated using methods that originate from participatory design, or co-design, which encourages the involvement of a wide variety of users or user groups in the design process. Such design approaches tend to rely on activities that take place in face-to-face contexts (Mahmud *et al*., 2020) and in which users are often heavily involved through multiple interaction occasions for articulating design needs (Islind *et al*., 2019a). However, in certain situations, user involvement can be difficult and, in extreme cases, even impossible. For instance, this was the case during the COVID-19 pandemic, when some users were unreachable for the face-to-face interactions needed to complete the designs. In such situations, when user groups are too fragile or unable to represent themselves, or when boundaries of some sort render it difficult to include them, it becomes critical to look for novel ways of involving these user groups in the design process to achieve high-quality designs.

During recent global events, it was virtually impossible to engage users in on-site activities because face-to-face interactions were deemed unsafe. When participants cannot join in person, one common option is to turn to online settings (Greenhalgh *et al*., 2020; Mastaneh and Mouseli, 2020). However, although online design activities can be valuable, some activities are not suited for online settings, and online participatory design settings have proved to be problematic and challenging (Bakırlıoğlu *et al*., 2020). Therefore, drawing exclusively on online design activities can be difficult and not feasible, particularly when users are fragile (Cerna and Müller, 2021; Näkki and Antikainen, 2008). Moreover, it has been shown that for online design activities to be a viable option, trust-building is needed, which established through face-to-face interactions prior to the online design activities, enables fruitful co-design sessions in hybrid settings (Islind *et al*., 2019a). Furthermore, scholars have argued that an important aspect of successful co-design is the need to engage in authentic settings established by working in design spaces where users can feel relaxed (Islind and Lundh Snis, 2018). Consequently, when users are unreachable or unable to participate and when online design activities are not feasible, alternative approaches to understanding users' needs are needed.

In addition, when several user groups are involved in the design process, they tend to be heterogeneous in terms of skills, experiences, and use contexts. Therefore, these groups can have radically different needs depending on how the digital infrastructure influences their everyday lives. The literature on participatory design and co-design describes ways of handling design with heterogeneous users—namely, through the iterative involvement of various user groups as early as possible and throughout the design process (Islind, 2018; Sanders and Stappers, 2008). However, this can be demanding for both users and designers because co-design sessions can be taxing, and the entire co-design process can last for a long time (Read *et al*., 2022). Thus, when initiating the design process, deriving multiple perspectives in a time-efficient manner is important as time is a valuable resource. More specifically, users' time should be respected, and in addition to that, some design processes involve a short timeline, which may require rapid advancement. Therefore, given that moving forward in a time-efficient manner is important for users and designers, getting to know users and understanding their needs through extensive interviews, co-design workshops and similar activities may not always be the most effective solution. For example, Joshi and Bratteteig (2016) reported finding an alternative approach to participatory design to enable the inclusion of a user group consisting of older adults because representatives from this user group were unable to attend every participatory design meeting. In sum, it is key to consider time and extensive involvement constraints in relation to both users and designers.

In this article, we consider ways of tackling the following two dimensions of the design process simultaneously: (i) the need for alternative ways of including users who are unable to represent themselves and (ii) the need to do so in a time-efficient manner. We report on a design project whose aim was to develop a digital infrastructure for supporting cancer patients who had undergone treatment and were living with long-term effects brought on by the illness and the subsequent treatment while also supporting the healthcare professionals caring for them. The clinic worked mainly with female patients who had survived cancer of the female reproductive system, bladder cancer, or colorectal cancer. The patients were dealing with issues such as urine and fecal leakage, sexual dysfunction, lymphedema, pain, and various urinary tract symptoms. This particular patient group had great difficulties visiting the clinic to be closely involved in design activities during our co-design sessions due to complex issues that rendered many of them unable or less able to leave their homes. In the process, the designers and researchers collaborated closely with the nurses at the clinic, who were greatly invested in the well-being of the patient users and highly knowledgeable about the patients' everyday struggles. Given that the patients were difficult to include and that some were even unable to participate due to the severity of their illnesses, it was important to consider alternative strategies for implementing a successful co-design process, such as engaging users in the design process to speak on behalf of those who were unable to participate. Moreover, as the patients needed to preserve their strength for survival and recovery, we deemed it kinder and more ethical to shelter them as much as possible. When searching for alternative user involvement strategies, the literature on participatory design and co-design provides few answers on what to do when users are unavailable. More specifically, there is a gap in the literature regarding the engagement of proxy users as an embedded element in co-design. Therefore, we posed the following research question: *How can roleplaying be embedded in co-design to engage users as proxies on behalf of those who are unable to represent themselves?* To address this research question, we conducted a design ethnography based on the design of a digital infrastructure at a cancer rehabilitation clinic.

Our main contribution to the literature is the conceptualization of proxy design as a method that can ignite and initiate the co-design process when important users are unreachable or unable to represent themselves. More specifically, based on empirical findings from a design ethnography that involved nurses as proxy users speaking on behalf of patients, we show that roleplaying is a fitting way of initiating the design process outlining proxy design as an embedded element of co-design. Consequently, we claim that the proxy design method should be incorporated into co-design. Finally, we outline six design guidelines that should be considered when engaging in proxy design.

The rest of the article is organized as follows: In the following section, we summarize related research on co-design and proxies, followed by proxy enactment through the use of scenarios. Then, we outline our methodological approach (i.e., design ethnography) for gathering the empirical data and present the findings from the longitudinal design process. Finally, we discuss our findings in relation to the literature and offer concluding remarks.

# 2. Related Work

As our focus in this article is on involving proxy users in co-design, and given that particular involvement was enacted through roleplaying sessions, this section outlines the literature related to these aspects. First, we discuss the state of the art of co-design, before considering the involvement of proxy users in design. Then, we summarize the literature on how roleplaying has been used in research. It is relevant to point out that the design process outlined herein included the design, development, and testing processes of the entire digital infrastructure rather than simply involving the "shaping" of a digital artifact. In other words, our design process included mapping and analysis of user needs, development, and iterative evaluations.

## 2.1. Co-Design and Proxies in Design

When participatory design, later named collaborative design and ultimately termed co-design, emerged in the literature, related design approaches were often geared toward designing a specific service. Since then, the situation has changed. Today, we are not only designing products for, or with, users; instead, we are designing complete experiences and digital artifacts in digital infrastructures that construct cultures and new practices (Farshchian *et al.*, 2021; Islind, 2018, 2022; Kempton, 2022; Willermark and Islind, 2022). Co-design goes beyond the mere involvement of relevant stakeholders, who will later become the users of the digital artifact. First, it is necessary to involve users early in the design process so that they can participate in the early development decisions. Second, as users are generally not trained in design prior to participating in co-design processes, it is necessary to allow the design space to become a learning opportunity in which users can reflect on their needs (Cerna *et al.*, 2020; Islind and Norström, 2020; Vallo Hult *et al.*, 2020). Finally, as the designed digital artifacts are often novel, it is necessary to foster users' abilities to imagine what digital artifacts will allow them to do in the future. Therefore, the fundamentals of the co-design approach entail providing users with the necessary skills that will enable them to participate fully in the design process. Consequently, users can have a say in the design processes that ultimately affect their lives and work (Joshi *et al.*, 2016; Kensing and Greenbaum, 2013). To sum up, co-design is a collaborative, democratic creative activity in which users who are not previously trained in design enactment engage with designers and one another to further the design process. Thus, co-design can be seen as an instance of co-creation (Islind, 2018, 2022; Sanders *et al.*, 2008).

As outlined previously in this article, in certain design processes, the involvement of specific user groups is not feasible. In such situations, the inclusion of proxy users can be a solution. To consider this option further, we now elaborate on proxy users in design. There are various motives for engaging with proxies rather than actual users—for example, when users in healthcare settings cannot participate due to personal issues (e.g., De Leo and Leroy, 2008; Guerrier *et al.*, 2019). In this context, a central element is the competence and knowledge of the proxy regarding the actual user group. For example, working with children with severe autism and communication difficulties, De Leo and Leroy (2008) involved their teachers in the design process as proxies, a group that was professionally trained to support these children and had actual experiences of their lives. Boyd-Graber *et al.* (2006), when developing desktop personal digital assistant systems for supporting patients with aphasia, recruited speech-language pathologists, a group with specific competencies in relation to the challenges of the targeted users. Furthermore, the relevance of proxies may differ depending on the targeted users as well as what is being designed. Ideally, proxies should have knowledge of and experience with both areas—see, for example, how Feng *et al.* (2010) involved the family members of children with Down's syndrome and how Cohene *et al.* (2005) involved caregivers as proxies for individuals dealing with Alzheimer's disease. As an alternative to involving proxies with actual competences and experiences related to the targeted users' conditions, some scholars have proposed relying on simulations with proxies. This means constructing use situations to acquire input for the design. For example, aiming to create an experience of the uncontrollable movements of patients suffering from cerebral palsy, Guerrier *et al.*

(2019) used simulations with proxies, whereby a rope was tied to the wrist of a healthy subject and then pulled to simulate uncontrollable movements, to conduct an efficient user study.

However, even when proxies are knowledgeable, they may face challenges in understanding and acting as representatives of the bodily experiences of actual users because they fundamentally lack the lived experiences of the user group. Similarly, Feng *et al.* (2005) argued that users with spinal cord injuries cannot be represented by people without spinal cord injuries. This applies even when people with and without spinal cord injuries have similar speech-related abilities (when testing speech-quality technology); they simply do not comprehend the severity of having a permanent spinal cord injury (Feng *et al.*, 2005). Feng *et al.* (2005) further argued that users often compare new digital artifacts that are being designed to previously used digital artifacts. Consequently, the lack of insight into other digital artifacts or assistive technologies used exclusively by people living with specific impairments, such as spinal cord injuries, can derail the design process. Nonetheless, design by proxy is appropriate in certain situations. For instance, proxy design can be fruitful when the individuals involved have in-depth insights into users' lives and are familiar with the detailed aspects of users' impairments (Lazar *et al.*, 2017).

Based on the studies that have been done with proxy users, it appears that proxies can vary from family members and teachers to caregivers. Moffatt (2004) and Davies *et al.* (2003) used advocate users and argued for advocate users' involvement because the latter are highly equipped to deal with the situation at hand. Furthermore, Moffatt (2004) and Davies *et al.* (2003) supported the involvement of people who are dealing with similar function variations or who have the same disease but of a less severe kind. Based on these arguments, we could have chosen patients who had not yet finished their cancer treatment. However, given that patients' needs shift over time and that early-treatment patients may be unable to verbalize the needs of those in aftercare, this was not a viable option.

## 2.2. Proxy Enactment through Roleplay

Adopting a role other than the one that a person has in their personal or professional context is a method that has been used in research for dozens of years (e.g., the famous experiments by Milgram [1963] or Zimbardo *et al.* [1971]). In design research, two types of roleplay are used. The first involves activities in which people play different roles (e.g., games or theatre). This is often the case with games (Domínguez *et al.*, 2016) and activities involving gamification elements (Huang *et al.*, 2020) when designing digital artifacts. The second involves designing activities that simulate other, original activities (e.g., a clinician talking with a patient, which is similar to what we used in our research). Such simulations can be conducted in various ways—for instance, to improve the competence and understanding of designers, the designers themselves can take part in simulated scenarios. This can be done, for example, by involving designers in scenarios that can increase their empathy in a particular type of situation, thus enabling them to more vividly imagine the experiences of future users (Wright and McCarthy, 2008). Another approach entails organized roleplaying scenarios to help the design team understand complex social concepts (Waern *et al.*, 2020). In co-design, parts of newly designed services are presented through scenarios played out by research teams (Sustar *et al.*, 2013).

Other approaches draw on roleplay to evaluate user experience in meaningful ways by involving users in scenarios. In one paper, roleplaying workshops were organized to evaluate the usability, acceptability, and feasibility of a digital infrastructure that was aimed at supporting nurses in their work and relocate patients' responsibility (Klemets and De Moor, 2015). The nurses participated in the roleplaying scenarios and acted as patients (Klemets *et al.*, 2015). West *et al.* (2016) used roleplaying to elicit feedback from future users. In a study of clinicians and their needs for self-tracking data in their work, the clinicians were introduced to scenarios based on real clinical cases and faced the task of giving advice regarding these cases (West *et al*., 2016). The purpose of that study was to understand which requirements are necessary for self-tracking applications that could be used in healthcare (West *et al*., 2016). In another study, users participated in scenarios to help the designers design a digital artifact while becoming, through scenario enactment, more aware of the possible risks connected to technology use (Downs *et al*., 2007). To sum up, roleplaying can be used at different stages of the design process. Such research activities involve acting out imaginary scenarios, often based on prototypes that the designers build according to the collected data. These scenarios can be directed at either designers or users, but they are developed by researchers or designers. We argue that during the design process, it could be worthwhile to explore simulated situations based on users' and proxy users' lived or professional experiences. This is particularly relevant when users are unavailable for observation or participation in the design process. Moreover, there is a gap in the literature regarding proxies as an embedded element in co-design.

Although the literature clearly states that user involvement is a vital aspect of the design process for establishing user-centered digital infrastructures, we argue that the proxy design method can be a valuable resource for those aiming to create designs for user groups that are unable to participate in longitudinal or extensive data collection processes. Proxy design is not meant to replace user involvement; instead, this method can be used as an alternative route in the design process when particular user groups cannot participate. In addition, we outline proxy design as an embedded part of co-design. To illustrate this, we will now discuss the data gathered and analyzed in this article.

# 3. Research Approach

When studying the design of information systems, there is a longstanding ethnographical tradition of studying and approaching the design process (Anderson, 1994; Beyer and Holtzblatt, 1999; Dourish, 2006). The aim of design ethnography is to explore shared design experiences to facilitate learning about social and cultural practices and values based on empirical cases. Design ethnography is a research approach in which the researcher actively engages with others regarding the future-oriented objectives of designing, developing, and improving artifacts that may affect cultural and social settings (Baskerville and Myers, 2015). In addition, design ethnography is not conducted before the design process begins; instead, the design process constitutes the ethnographic field, and researchers conduct participatory observation within the design project (Baskerville *et al*., 2015). In contrast to traditional ethnography, in design ethnography, researchers play an active role and are engaged in the process of designing artifacts; this was also the case for our research initiative. Because the design process and the proxy design method constitute the unit of analysis in this article, design ethnography was the guiding methodological thread of our study. Although the generation of a novel artifact is an important step in design ethnography methodology, it is not the main contribution of the research process. In our case, several new artifacts were produced, but the goal of this article is to report on the initial phases of the design process and conceptualize proxy design as a method. In other words, the contribution of design ethnography is in-depth knowledge about specific design experiences in particular practices, about stakeholders, goals, and action spaces, and about ways of turning this particular knowledge into more generalizable design knowledge (Baskerville *et al*., 2015). In our design ethnography, we followed the same objective.

We applied the co-design method to the entire design process. More specifically, this approach involved proxies in simulated and roleplayed consultations between nurses and patients in which the nurses played both roles (one nurse per role per roleplaying session). The motivation for involving the nurses as proxies was that nurses spend time with patients daily over the years, understand the complexity of patients' problems, and have extensive experience with particular groups of patients. Drawing on their experiences, the nurses roleplayed patient–nurse consultations. As stated earlier, there were several reasons for this method choice: (i) we wanted to shelter the patients, be mindful of the time of the users and designers, and we deemed it kinder and more ethical to limit the early patients' involvement; (ii) the patients were difficult to reach; and (iii) involving "all types" of patients would have required approximately 15 individual patient interviews. Therefore, the reasoning was rooted in efficiency for everyone involved. The proxy design method involved three simulated consultations in each three-hour workshop (nine simulated consultations over nine hours in total). Furthermore, the empirical data were based on our involvement in the project for three years after the workshops took place. We built an extensive digital infrastructure based on these initial roleplaying sessions and the proxy design method, which was embedded in a larger co-design process. Our findings are derived from this entire process.

The roleplaying sessions took place at the Sahlgrenska University Hospital in Sweden where the clinic was located. The nurses roleplayed patient–nurse consultations: one nurse played a patient, and another nurse acted as a nurse. The proxy method involved three simulated consultations in each three-hour roleplaying workshop. This setup meant that we conducted a hybrid form of design experiment and design ethnography. More specifically, by recreating and enacting use situations and engaging with these situations using our design ethnographic method, we could explore the use situations in an efficient and user-centered fashion. The design and development focused on the following three main digital artifacts: (i) an information portal, (ii) a mobile app, and (iii) a backend accessible through a digital platform (which included a video-based consultation tool and visualizations of patient data, further discussed in Islind *et al.*, 2019b). The combination of these three artifacts is referred to as 'the digital infrastructure.' The first artifact provides accessible information to patients. The second is a data-collection app in which patients can gather data. The third artifact provides a way for nurses to view patient-generated health data to support their decision-making processes and have video-based consultations with the patients. Thus, the app and the digital platform include patients' personal data that can be viewed by nurses.

The entire project lasted over three years (the digital infrastructure is now in use), and the empirical material consists of much more than the data presented in this particular article (see Islind, 2018; Islind *et al.*, 2019a, 2019b; Cerna *et al.*, 2020). In total, there were three roleplaying workshops (also called proxy design sessions) that established the basis for the empirical data presented in this article. Each workshop lasted three hours. The empirical data analyzed in this article are presented in Table 1. During the proxy design workshops (see Table 1), two researchers were present (who were also designers and are referred to as researchers and designers interchangeably), alongside a communicator who helped facilitate the workshops. Moreover, the empirical data presented in this article also include three co-design workshops that lasted two hours each, with three to four patients and two or three nurses present during every workshop. These workshops took place after the proxy design sessions and will be discussed in the Findings section alongside the proxy design sessions. In addition, the empirical data included an observation study and recordings of telephone consultations between actual patients and nurses. Therefore, the results presented in the Findings section are based on both observation notes and verbatim transcripts. The data were translated from Swedish to English for the purposes of this article.

| Activity | Description | Objective and data gathered | Duration and participants |
| --- | --- | --- | --- |
| **Proxy design sessions** | The proxy design workshops were conducted to initiate the co-design process because the patients were unreachable at the beginning | The objective was to initiate the design process<br><br>The data gathered were audio recordings, transcribed verbatim for each workshop. Each workshop included three simulated consultations | 3 workshops in total, each lasting 3 hours, 9 hours in total<br><br>Present at each were 3 nurses, 2 researchers, and 1 communicator |
| **Observation study** | Observation of the work at the clinic; the observations included face-to-face patient consultations and daily work at the clinic | The objective was to understand the clinical practice and the interactions with patients<br><br>The data gathered were field notes from the observation study, which informed the entire co-design process | 20 full workdays of observations at the clinic<br><br>Present was a researcher and 3 nurses; on some days, a physician and patients were presented |
| **Co-design workshops with nurses and patients** | Workshops for bringing together patients and nurses to evaluate the digital infrastructure | The objective was to involve patients alongside nurses<br><br>The data gathered were audio recording, transcribed verbatim, as well as field notes from each workshop | 3 workshops in total, each lasting 2 hours, 6 hours in total<br><br>Present at each workshop were 3–4 patients, at least 2 nurses, and 2 researchers |
| **Telephone consultations** | Telephone consultations between patients and nurses | The objective was to gain in-depth insights into how nurses and patients used and talked about data and the digital artifacts<br><br>The data gathered were audio recordings, transcribed verbatim, for all telephone consultations | 21 telephone consultations, 15-50 minutes each, 9 hours in total<br><br>Present at each consultation was a patient, a nurse, and a researcher |
| **Log data from the mobile app** | Log data from the patients' use of the mobile app; aggregated and used to understand the use behavior | The objective was to gain insights into the use of the mobile app<br><br>The data gathered were log data, containing click-logs from the mobile app, used to shed light on the use of the digital artifact (the app) | Log data for a year of each click in the mobile app made by patients and used to understand their clicks when using the mobile app |

*Table 1. Data-gathering activities. Source: Table created by the authors.*

The nurse-led clinic consisted of three oncology nurses, specially trained in understanding and working with issues concerning cancer survivorship and rehabilitation, with a focus on self-care strategies and treatments. One of the most troublesome symptoms from which patients suffer as cancer survivors with a history of pelvic radiotherapy is frequent bowel movements and fecal leakage. Sexual dysfunction, lymphedema, and urinary-tract symptoms are long-term late effects discussed by patients and nurses. In the roleplay sessions, all three nurses took turns playing a patient and a nurse, drawing on the cumulative knowledge of these symptoms.

The analysis was conducted using content analysis (Graneheim and Lundman, 2004) and abductive engagement with the empirical material in the context of longitudinal engagement with design ethnography. Here, abductive engagement refers to the interplay between the empirical data, gathered in relation to real-world problems (inductively obtained), and theory (deductively inferred), established by viewing "reality from the theoretical viewpoint or perspective" (Van de Ven, 2007, p. 104). In other words, abductive engagement involves shifting between inductive and deductive reasoning to continuously revise, sharpen, and reformulate the research design (Gregory and Muntermann, 2011; Van de Ven, 2007) based on the analysis of the empirical data. The coding of the empirical data involved three rounds conducted by the first author aimed at narrowing down the analysis, presenting intermittent findings to the co-authors of this article, and iteratively discussing those in the larger author group. The following section presents the findings derived from the empirical data.

# 4. Findings

## 4.1. Simulated Scenarios

The first design meetings were shared sessions between the nurses, a patient representative, and the researchers. However, as the project progressed, the patient representative was unable to participate, and recruiting new patients turned out to be difficult due to their extensive health-related problems triggered by cancer and its subsequent treatment. We started mapping out the needs of the nurses, but the patient perspective was missing. To speed up the process while sheltering the patients, we decided to set up roleplaying workshops with the nurses to act out the consultation scenario. The first roleplaying workshop involved three personas based on the insights into the patients who suffered from serious bowel problems. One persona suffered from urgency issues, another from leakage, and the third from gas. During the second roleplaying workshop (see Figure 1), the nurses discussed introducing three personas for this session as well. The problematic aspect that was the focal point was sexual problems. The nurses discussed the specifics of each persona and how the sexual problems of the patients could be conveyed through the personas. They were in consensus regarding the importance of establishing the right disease diagnosis before the start of the roleplaying workshop, as illustrated in the following excerpt from transcripts from the second roleplaying session:

> Nurse 1: I have thought about the diagnoses. One with rectal cancer, one with anal cancer, and one with some form of gynecological cancer.
> Nurse 2: Yes, cervix or maybe corpus cancer?
> Nurse 1: Yes, maybe cervix. Anal, rectal, and cervix [cancer]. Then, we cover all our problems because the cervix patients get the toughest treatment of the gynecological patients, so we'll really cover it [the widest spectrum of patients' problems].
> Nurse 2: Yes. Just that the cervix has no, or well, they have not had surgery.
> Nurse 1: Well, but it doesn't matter, I do not think so.
> Nurse 2: No.

> Nurse 1: We may not need to cover everything. Well, rectal [cancer] patients have had surgery, but anal [cancer] patients have not had surgery.
> Nurse 2: I just think that corpus [cancer patients] is quite a big group. Cervix and anal are the most similar.
> Nurse 1: Well then, maybe we can remove the cervix patient [from the discussion], and then we take the corpus [patient] instead? Because they have had a relatively similar radiation dose.
> Nurse 2: Anal has other problems. Do we have to make three personas, or how can we do this?
> Designer: There is nothing that says there must be three [we did three last time]. If we cannot get it down to three, and we feel the need for more than three, it can definitely be four or more.
> Nurse 2: A large group is, as I see it, endometrial, anal, and rectal in numbers and people, but cervical cancer patients have the most problems. Can we talk about desire problems? Menopause problems? Hormonal problems? Radiation-induced problems?
> Nurse 1: Or should we talk about it from the patient [perspective]. Like, think what a patient has been through, and what the patient has for problems? Or how do we do this best?

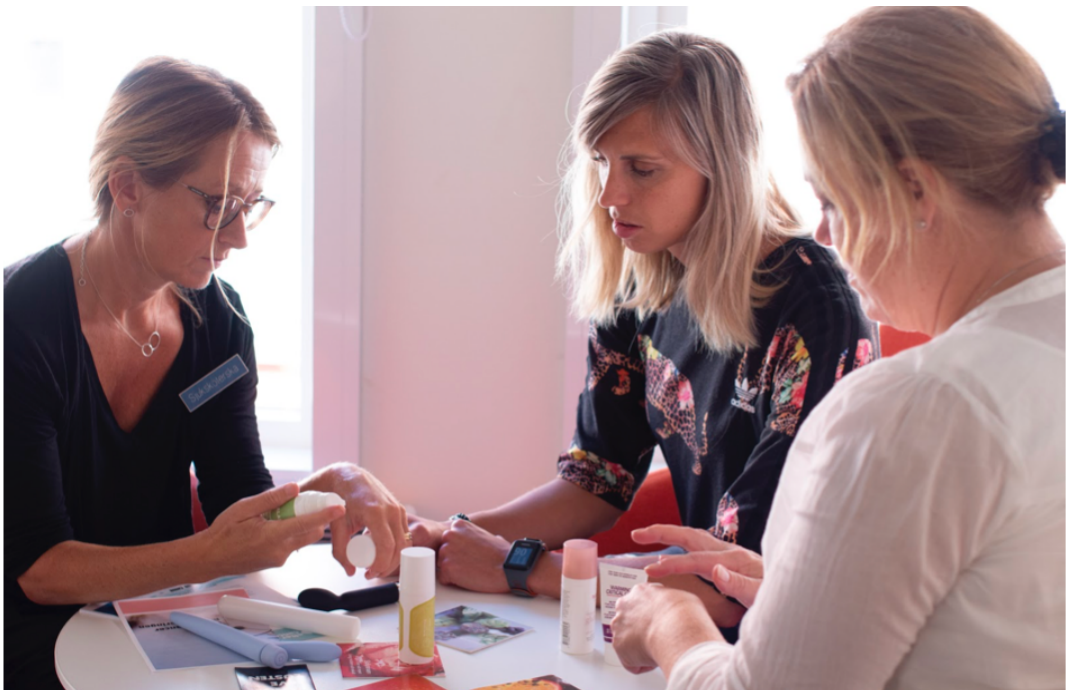

*Figure 1. The nurses in action. Source: Author's own photo.*

The nurses pinpointed the core of the issue of patient involvement. The patients came from a wide spectrum of cancer diagnoses and treatment types and were dealing with a range of problematic issues. For the team to establish the basis for the future digital infrastructure supporting nurses' work while also supporting patients' self-care, it was important to find a balance between personalization through data and scalability. Through a careful negotiation of the different diagnoses and their characteristics, the nurses started delimiting the features that should be included (patient number, radiation intensity, and essential data collected) to limit the vast range of patients' experiences (instead of interviewing representatives for all particular diagnoses). The nurses continued to discuss the heterogeneity in the patient group and the wide range of sexual problems with which the latter dealt. We continued by simulating consultations with patients through roleplaying. The roleplaying was meant to represent the nurses' clinical experiences of meeting a range of patients for each diagnosis type. The second roleplaying workshop also involved a set of personas dealing with sexual problems. The nurses decided on the diagnosis for each simulated consultation through discussion and brought the symptoms to discussion prior to instigating the roleplaying session. Their discussions were not merely based on a "typical" patient; instead, the discussions included elements from a wide variety of patients suffering from rare to frequent issues. This approach addressed the aforementioned balance between personalization and scalability, allowing for the inclusion of a large number of patients' experiences while also covering issues that were both typical and uncommon. One nurse played a patient, and another nurse played a nurse during the roleplaying workshop with simulated consultations. The following is a description of a scenario captured from the session:

> The proxy patient (played by the nurse) enters the consultation to discuss their complex problems related to sexual issues caused by cancer and subsequent cancer treatment. The nurse starts off by asking questions related to patient history and says that she has read the EHR (electronic health records). The patient asks questions. The patient explains in detail a complex sequence of problems. The patient has extremely complex problems, and the range of the problems explained represents the experiences of patients who come to the clinic and have problems with almost everything. The problems described range from bladder control, bowel movement issues, and complex sexual problems, which are discussed in limited detail. The nurse asks follow-up questions to acquire more details regarding the limited aspects of the story.
>
> From the observation notes: [It is apparent that the nurses use coaxing tactics to extract the information from the patients. The nurse playing the nurse employs a sympathetic tone, nods to show compassion, and is professional when establishing facts about the issues at hand.]

During the roleplaying workshops, the nurses changed the roles, and the next patient had less comprehensive problems. The nurse who had played the nurse in the previous scenario played the patient in the following scenario. The exchange of roles was smooth. All three simulated consultations in all workshops included several patients combined that the nurses had in mind while describing the problem area. The third simulation in each roleplaying workshop included a patient who was further along in the treatment at the clinic. The patient in the third simulated consultation had complex problems, and the consultation simulation ended with the nurses deciding on a prioritized list of symptoms to tackle first. The list was to be iterated and updated during the following treatment. As in the first roleplaying workshop, in the second and third roleplaying workshops, we went through three simulated consultations, and the three nurses who were present took turns playing the patient and the nurse. The second roleplaying workshop resulted in personas dealing with sexual problems (see Figure 2).

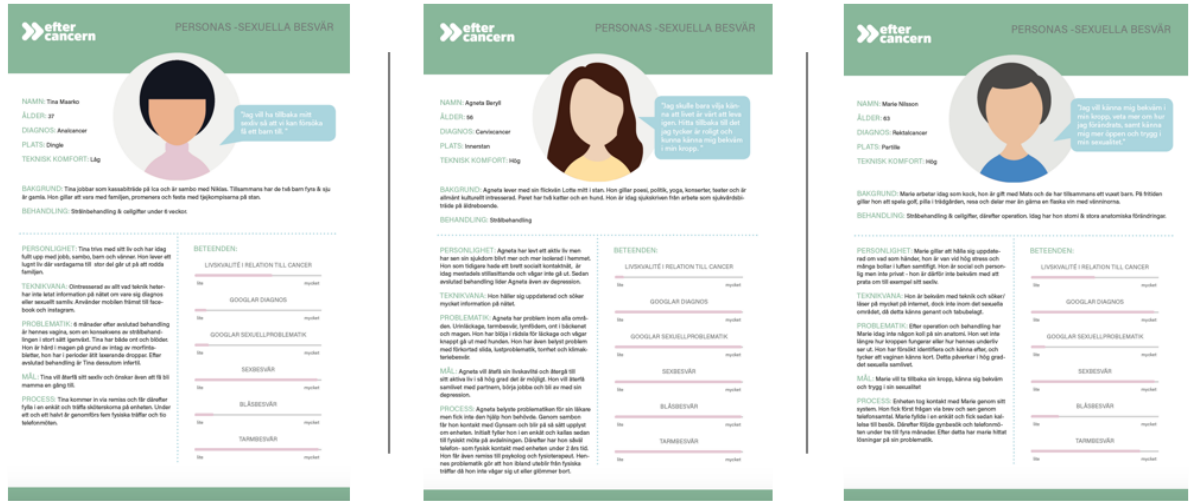

*Figure 2. Patient personas abstracted from the simulated consultations from the roleplaying workshops. Source: Source: Personas created by the authors.*

These personas constituted the basis of the users' stories that were developed in the third roleplaying workshop. Moreover, we used the user stories, along with the personas, to initiate the design process. A week after the second session, we met again and had a similar simulation of three consultations based on patients with bladder-related issues to develop user stories. The following excerpt comes from the third roleplaying session. In the excerpt, the quotations labeled Nurse 1 refer to the nurse who was roleplaying the patient, while Nurse 2 roleplayed the nurse:

> Nurse 1: So, the problem is when it leaks. I have to use a pad.
> Nurse 2: Is the pad type that you use enough, or do you think you need something else, maybe?
> Nurse 1: Yes, most of the time, it is sufficient.
> Nurse 2: But the leakage is a problem. I understand that. Of course. We can maybe try this [urination assessment table] that we have. Then, you can fill it in and follow how much fluid you take in, and how much comes out, would you like to try that?
> Nurse 1: Well, maybe. If you think it could help. The problem is sometimes leaking.
> Nurse 2: Maybe we need to know how big of a problem it is.

In this excerpt, the nurses are discussing the options for a patient who has a leakage problem. One nurse proposes using a urine assessment table, which is the clinical standard for collecting data on urine leakage problems. It also represents the types of tools that nurses use in their practice—in particular, they see the urine assessment table as a tool for supporting patient learning and decision-making through data collection. More specifically, for nurses, the table is a tool that produces data they can work with and make decisions about patients' treatments. However, for patients, the table is a learning tool because it assists with self-care. This is essential to understand because digital artifacts for capturing patients' experiences need to match this context. We ended up including this in the mobile app and the digital infrastructure, which were premised on the described design process.

## 4.2. Prototype Evaluation with Patients

The roleplaying workshops were followed by co-design sessions with a focus on revising the developed personas and user stories with patients, a phase that continued until there was a consensus that sufficient materials were gathered for the prototyping phase. Initially, the prototypes were wireframes that, over time, grew into interactive digital prototypes. The digital artifacts were as follows: (i) an information portal, (ii) a mobile app, and (iii) a backend accessible through a digital platform (which included a video-based consultation tool and visualizations of patient data, as further discussed in Islind *et al.*, 2019b). As mentioned earlier, we refer to this combination of digital artifacts as *'the digital infrastructure.'*

We are aware that we relied heavily on the three nurses who acted as proxies and that this impacted the design of the digital artifacts. In most co-design processes, the outcome is dependent on the users who participate in the co-design process, and the same was true in our case. We addressed this issue by testing the digital infrastructure with patients. Therefore, after the three roleplaying workshops, patients were engaged more strategically through interviews (analyzed in a different article [blinded reference]) and workshops. The discussion shown below comes from a co-design workshop with three patients and two nurses whose aim was to discuss the digital infrastructure. It is important to note that the patients who participated in this workshop had access to the information portal, were asked to use a prototype of the mobile app for two weeks prior to the co-design workshop, and had participated in video-based consultations via the digital platform. The co-design workshops functioned as a forum in which discussions occurred between the patients, the nurses, and the designers; thus, engagement was not conducted individually. The co-design workshops were semi-structured, following a workshop guide but open to free speech. The following conversation comes from a discussion about the information portal:

> Patient 1: It is very clear and easy to understand. Definitely. I was like, "I did not know that when I came into the site. That was the smart thing for me." Then, I forgot to think about what it looks like and so on.
> Patient 2: I think it works very well indeed.
> Patient 3: Well. A lot has happened in just one year [since treatment started, they are dealing with the aftermath of their cancer treatment]. Really great to have information. Some parts are not developed thoroughly, but it is developing all the time [...] I'm just a little, like, thinking, about the possible extensions that could help me, now.
> Nurse 1: You really don't just have to say that it's good [everyone laughs]. You can say, change colors, I miss this, change this text, and so on. Like negative aspects. Can you imagine that this can support family members too?
> Patient 1: Yes, definitely, it's often so much information [in consultations], so it's hard to take everything in. It's good to give them a hint that they can go in and read. Go in there. Well, I have an example. My ex did not take in one third of what was told. My brother knows more about my cancer than my ex did. The closest relatives can block themselves off, and then they are not with you on the journey at all later on, so it can be difficult. It's great that they can also go in and read the same information as we.
> Patient 2: It is actually a good idea to ask them to read. I myself have explained and so on. Sure, it's really good if they read too, it's so easy if everyone shares the information. The site is very easy to understand. Definitely.

This excerpt illustrates that the patients perceived the portal as informative and considered it potentially useful for their loved ones. The discussion continued for a while, turning to the mobile app. The patients, who belonged to different age groups and had different types of cancer (thus dealing with different types of problems), had been using the prototype of the mobile app for some time (most of them two weeks prior to the workshops):

Patient 3: You do not want to click and click and click. It's better if it's more combined [the questions in the daily form]. I think it's the same as when you receive an information letter. Then it's more complete with questions, you don't get one question per page.

Patient 1: I think it's much better to have a question on one page. Like yesterday, I sat with my son, and he was registering for the military. For his part, it is even easier to answer one question per page [on an online form]. For me, it's also better to have a question on each page. I want that.

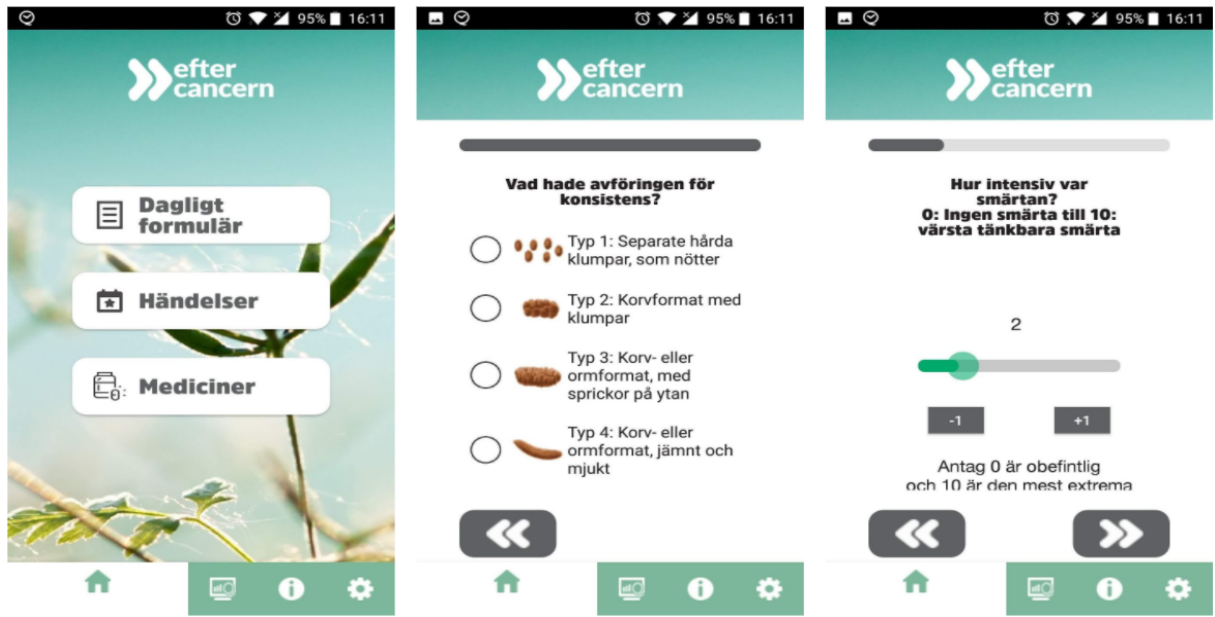

*Figure 3. Illustrations from the daily questionnaire in the mobile app. Source: Screens created and owned by the authors.*

Regarding the mobile app, we grouped some questions in the mobile app, while other questions required a button click to get to the next question in the daily form (see Figure 3). The excerpt above illustrates well that the patients were a varied group not only in terms of different needs connected to their health problems but also regarding their digital preferences. They continued discussing and concluded that one question per page was a viable option (which became the ultimate design decision). They stated that they learned the order of the daily questions after a while and could quickly fill in the form. The discussion led to other parts of the mobile app design, as mentioned by Patient 3: "This with the minutes is not good." They all nodded, expressing agreement. The patient was referring to a function for measuring specific incidents (pain or daily activity) in terms of time; that function was removed following this co-design workshop. The patients all agreed that the measurements were difficult to grasp, and while discussing negative aspects, they also described medication-related issues in the mobile app. Some medications were difficult to measure in grams, which was a hindrance. The patients could find workarounds and, instead, skipped the registration of the medications. Therefore, the medication part of the mobile app was redesigned and revised after this co-design workshop.

## 4.3. Contextualizing Digital Artifacts

After the co-design workshops (and several other design-related activities), the digital artifacts progressed from prototypes to digital artifacts tailored to support nurses' work and the daily lives of patients. This tailoring was a slow, iterative process of using and redesigning the digital infrastructure. For instance, the mobile app was used to collect patient-generated health data as a part of the treatment process at the clinic and to support the nurses' data-driven decision-making.

During the telephone consultations during which the mobile app was used as a support tool, its interaction design and usability characteristics were frequently brought up. After the patients started using the mobile app on a daily basis, during the consultations with the nurses, they highlighted elements for improving the mobile app:

> Patient: Can I just say one thing about that [the mobile application]?
> Nurse: Absolutely.
> Patient: When I started using the app, then Number 3 and Number 4 [on the Bristol scale] say almost exactly the same thing [the text], and I understood that it is not right.
> Nurse: No, exactly.
> Patient: It's almost exactly the same text, like: soft and formed as a snake.

Other aspects that came up during the consultations were related to the patients' own "practices" of using the mobile app in their own ways. They had their own tailored uses and informed the nurses about their ways of using the mobile app. For instance, one patient said the following during a consultation: "Sometimes, I have not had it [the smartphone] on me exactly when I go to the toilet, but then maybe I fill in three times [activities] at once."

The activities referred to by the patient were the registrations of bodily functions in the mobile app. More specifically, the patients had their own phones on which they had downloaded the mobile app. When the patients used the app, they reported each toilet visit, medication intake, or pain occurrence, and registered specifics directly in relation to the activity. Consequently, the patient fills in medication usage, along with each 'activity,' such as pain instance, toilet visit, or leakage. In this sense, their uses of the mobile app varied. The patients also fill in a daily questionnaire assessing the day.

From the nurses' perspective, there were several impactful findings. Prior to the design process, the nurses identified problems with collecting precise data on paper regarding toilet habits. Moreover, it is known that verbally collected patient anamnesis is affected by memory bias, which further strengthened the use of the data from the mobile app. However, prior to the design process, there were doubts about the benefits of employing digital tools in the clinical setting, but they vanished over time. The nurses saw the enormous advantages of using the data from the mobile application as an integrated part of the nurse–patient consultations. First, the use of the mobile application was found to be beneficial for increasing patients' understanding of their symptoms. It enabled visualizations of self-assessed data for a specific period both in the app and the digital platform. Based on the data, nurses could make more precise and data-driven decisions. Second, the documentation of patients' toilet habits proved to be more precise than verbal or paper-based data collection. Moreover, the participants also reported that the digital infrastructure improved patients' self-management capacities, enhanced their compliance and helped them navigate their new situations as cancer survivors in a data-driven manner. Furthermore, video consultations using the digital platform were seen as beneficial, primarily due to geographical factors, i.e., the patients could stay home while attending the consultations which enabled care for patients who lived far away in a more efficient manner. Finally, video consultations provided a richer experience for some patients than consultations at the clinic.

The log data from the mobile app use showed that one registration of a toilet visit took eight seconds and eight clicks on average from the time that the mobile app was opened until it was closed. In the extracts above, the patients discussed the clicks in terms of how many clicks there "should" be. On average, a patient reported different types of data 4–10 times a day (according to the log data). Thus, even with many toilet visits during the day, the total time of using the mobile app in seconds was relatively low—less than two minutes for the average patient over the course of a day.

# 5. Discussion

As argued earlier in this article, co-design is about achieving a solution and balancing design decisions in relation to what is relevant to specific users and their contexts (Fischer, 2018). Due to the open or ill-defined role of the digital artifacts prior to instigating the design process, the roleplaying sessions provided the designers with an in-depth overview of the patients' health issues while also enabling the designers to understand the nurses and their practices. When the nurses were engaged as proxies to speak on behalf of the patients, they acted as patients in the simulated consultations, and they were challenged to focus on the patients' perspectives and familiarize themselves with the patients' situations. Furthermore, they exposed their levels of clinical knowledge and communication skills to the other nurses, which facilitated trust during the roleplay. Therefore, roleplaying during proxy design also contributed to collegial exchange, and the nurses learned from one another, which contributed to in-depth competence development (Vallo Hult *et al.*, 2020). The nurses stated that the experiences of participating in the design process and roleplay were beneficial to their understanding of the patients' lives. The collegial exchange and mutual competence development were facilitated on several levels, and the digital artifacts became a valuable complement to existing clinical methods. We will now discuss our findings according to the following three themes: (i) the challenge of bias in proxy design, (ii) proxy design as an embedded element in co-design, and (iii) six design guidelines that should be considered when engaging in proxy design.

Regarding how such a process can be handled as an embedded element in co-design, and to address the issue of who can act as a proxy user, our proxy design is a vital alternative in situations in which an ecology of digital artifacts needs to be designed simultaneously for work and everyday lives. We suggest that proxy users be involved from the very beginning and that the involved proxy users possess either in-depth insights acquired from patients or collective insights built up over time based on a variety of interactions. Our findings suggest that the method can become a tool for competence development; as illustrated by our case, the nurses learned about their own practices, understood the digital tools, and tried walking in the patients' shoes. The conceptualization of proxy design as a method, which is the main contribution of this article, helps to address the problems of personalization and participation overload (Fogli *et al.*, 2020). First, proxy design helps generate insights early in the design process with the help of knowledgeable users who can be involved to lessen the participation burden for vulnerable users. Second, proxy design helps balance personalization and the need for standardization in healthcare by limiting the broad complexity of patients' experiences to the relevant symptoms that need to be alleviated.

## 5.1. The Challenge of Bias in Proxy Design

In this article, we draw on design ethnography with the following two user groups: nurses and patients. Some of their needs overlapped, whereas others diverged. This finding is in line with that of Vassilakopoulou *et al*. (2019), who identified shared information spaces for coping with divergent interests between healthcare professionals and patients in design.

The involvement of nurses as proxies involves potential issues in terms of favored functionality, as the roleplaying sessions in the proxy-design part of the co-design process may have been biased toward the needs of the proxies over those of the proxied user group. However, although the design was ignited by and initiated with proxies, the value of the digital infrastructure could have been evaluated by the users (and not merely by the proxies). This is in line with Islind (2022), who argued for extensive evaluations by end users in co-design. Therefore, although the dual role of the nurses could have resulted in a potential bias toward their own needs, we argue that despite this risk, the proxy design method as an embedded element in co-design is immensely valuable. The role of the healthcare professional is, by nature, oriented toward making decisions, whereas patients are more problem-centric, which may create another potential bias. Moreover, the nurses could have potentially gone beyond what the patients might have preferred and needed, as their accumulated knowledge of patients over years of experience might have created the illusion that they knew better than the patients (Thorseng and Grisot, 2017). We argue that all of these issues can be tackled through user testing with the user group later on. Our findings show that as the digital infrastructure was extensively tested with the patients themselves later in the design process, the biases were not present. Nonetheless, these are valid concerns that need to be acknowledged when applying the proxy design method.

Although proxy studies are often dismissed as biased or as a way of circumventing relevant user involvement (De Leo and Leroy, 2008), we argue that this method has high potential in contexts such as the one described in this article. More specifically, we have outlined ways in which roleplaying can be used as an embedded element within co-design to engage users as proxies for those users who are unable to represent themselves. Our recommendation is that the proxy design method be used for kickstarting co-design projects, for sheltering vulnerable users while bringing in representatives from the actual users later on, and for user experience and usability testing.

Importantly, engaging in proxy design does not mean that the goal is not true user engagement or true co-design. When undertaking co-design with heterogeneous users, co-design should be conducted with involvement of the users as early as possible, with all stakeholder groups, and throughout the entire design process (Islind, 2018; Sanders *et al*., 2008). However, when this is not possible, or when a significant user group is unreachable, we propose involving proxy users to initiate the design process. Our findings support the use of the proxy method for kickstarting the design process in an effective way. The simulated patient consultations during the roleplaying sessions allowed the nurses to better understand the patients' perspectives. Understanding the patients' perspectives was an important aspect of the nurses' clinical practice, and it also facilitated the integration of the new digital artifacts. Furthermore, the nurses showed one another how they conducted their consultations. Although the nurses discussed patient cases regularly, they did not have a concrete idea of how their colleagues conducted consultations prior to the roleplaying sessions. By becoming involved as proxies, the nurses shared and reflected on their strategies for coaxing information from patients. In addition, by creatively engaging in forming the scenarios during the roleplaying sessions, the nurses could come up with relevant routines for the emerging digital artifacts, as space was created for using these artifacts during consultations. Although there are clear challenges of bias in proxy design that need to be carefully considered, we illustrate that proxy design is an impactful method for initiating design.

## 5.2. Proxy Design as an Embedded Element in Co-Design

The collaborative development of personas, use scenarios, and prototypes by the designers and proxies was clearly shaped by the proxy method. Generally, the aforementioned design activities are commonly used by trained designers, whereas in this design process, they were employed by novice designers (i.e., the nurses). The findings obtained during the proxy design facilitated the development of new skills, namely, design skills. Therefore, roleplaying can be considered a design activity that can help shelter vulnerable users from extensive exposure in the design process and that facilitates the cultivation of design skills with and by proxies. These findings are in line with those of Joshi and Bratteteig (2016), who reported that the inclusion of a user group consisting of older adults as representatives of the user group was not sensible because they were unable to attend all the design meetings. Moreover, we used this newfound design activity of roleplaying as an embedded element of the co-design process—that is, the proxy design method was employed in the early phases only and was followed by prototype testing with all users.

Regarding the qualities that can be introduced into the co-design process by proxy users when the actual users are unreachable, our design ethnography illustrates that the involvement of the nurses as proxies supported a resource-effective initiation of the design work. Although the patients were not involved in the co-design process from the very beginning, we could create a digital infrastructure that enabled the nurses to make sense of the patients' situations and test the digital artifacts with patients later in the co-design process. The proxy method enabled us to retrieve valuable information over a relatively short period of time. De Leo *et al*. (2008), Boyd-Graber *et al*. (2006), Feng *et al*. (2010), and Cohene *et al*. (2005) all highlight the importance of involving proxy users when the actual users are unable to express their thoughts or participate due to severe conditions. However, we propose that the involvement of proxy users can serve two additional purposes: (i) to effectively initiate the design process and (ii) to include users who are unable to participate either due to geographical issues or other types of boundaries. In such cases, our proxy method can be employed as an embedded element in an extended co-design process. Moreover, Moffatt (2004) and Davies *et al*. (2003) advocated the involvement of people who are dealing with similar functional variations but of a less severe kind; in other words, people who are dealing with the same disease but in a less severe form. However, given that our kind of users only tend to seek healthcare when their issues are strenuous, finding users with the same problems in less severe states proved difficult. Furthermore, the approach of involving the nurses had the added value of collegial exchange, which is a key benefit of involving proxy users in design in general and in co-design in particular. Although Feng *et al*. (2005), echoed by Lazar *et al*. (2017), argued that proxy users do not have the experience of actual users and thus cannot replace them, we found that the involvement of carefully selected proxy users with: (i) in-depth insights into actual users' histories; or (ii) collective insights obtained from working with a wide variety of heterogeneous users over an extended period of time can make proxy users highly valuable in proxy design settings.

## 5.3. Six Design Guidelines for Proxy Design

The notion of design with proxies has almost exclusively been elaborated with users who are unable to participate due to impairment, whereas the users in our study could participate verbally and cognitively but, due to physical symptoms and unpredictable bodily functions, suffered from impaired participation. Drawing on our cumulative findings, we conceptualize proxy design and outline the following six design guidelines for applying the proxy design method: (i) proxy design is a way of gaining an initial understanding of users to develop personas and user stories; (ii) proxy design is a way of collecting the information needed to design and develop the targeted digital artifacts; (iii) proxy design is a way of efficiently gaining an in-depth understanding of a user group that is complex, heterogeneous, has a wide range of problems, and is unreachable or unable to participate; (iv) proxy design is a way of starting the design process before involving the actual users (in addition to their proxies); (v) proxy design is a way of reducing participation overload for frail users; and (vi) proxy design is a way of facilitating collegial exchange for the proxies involved in the roleplaying sessions.

We will now elaborate on each of these points in turn. First, proxy design can be used as a way of gaining an initial understanding of users to build personas and user stories (or to engage in similar design activities). This guideline highlights the impact of proxy design on the start of the design process. Second, proxy design can be used as a way of collecting the information needed to design and develop digital artifacts. This guideline highlights the impact of proxy design on the gathering of quality data to support the design process. Third, proxy design can be used as a way of efficiently gaining an in-depth understanding of a user group that is complex, heterogeneous, has a wide range of problems, and is unreachable or unable to participate. This guideline highlights the impact of proxy design on the acquisition of an in-depth cumulative understanding of complex, heterogeneous user groups. Fourth, proxy design can be used as a way of initiating the design process before involving actual users (in addition to their proxies) in the testing process. This guideline highlights the impact of proxy design on the initiation process, as well as the importance of testing digital artifacts later with all user groups. Fifth, proxy design can be used as a way of reducing participation overload for frail users. This guideline highlights the potential of proxy design to decrease actual user involvement early on by drawing on cumulative knowledge of the proxies. This strategy can be valuable when testing with actual users is taxing and costly in terms of time and when the knowledge of the proxies is extensive. Sixth, proxy design can be used to facilitate collegial exchange and mutual competence development for the proxies involved in the roleplaying sessions. This guideline highlights the impact of proxy design on mutual competence development within the group that enacts the scenarios during the roleplaying activities. Finally, our findings suggest that an important balance can be established when involving proxy users through roleplaying activities early on in the co-design processes. More specifically, by creating a design space in which proxies can imagine what it is like to be part of the actual user group, the two different user roles can be synthesized into one.

## 5.4. Limitations and Future Work

The use of the proxy design method to initiate the co-design process involved certain limitations. For example, despite the nurses' vast knowledge of the patients' problems, the roleplay activities build solely on the accumulated experiences of the nurses. This can be particularly problematic when roleplay is used to emphasize what it means to be a certain person instead of focusing on which digital features need to be present for that person to use the digital infrastructure (Bennett and Rosner, 2019). Consequently, it is possible to overlook extreme or rare cases. Furthermore, there are methodological limitations to design ethnography, as this method takes time (thus, three years of design engagement) and limits the findings to a specific context, with the generalization of the results being difficult (Shapiro, 1994; Twidale *et al*., 2014). In addition, although this method allowed the nurses to learn about their new tools, the patients' learning through their involvement in the co-design process was postponed due to the use of the proxy design method. Although the data were limited in terms of sample size, the richness provided by design ethnography makes up for this shortcoming. Future work should consider exploring the proxy design method as an embedded part of co-design in other contexts and with larger sample sizes. Finally, it may be worthwhile to explore next-of-kin in proxy design.

# 6. Conclusion

The design of digital artifacts, the ecology of digital tools, or any digital infrastructure cannot be seen as a linear, straightforward process. This is particularly the case when dealing with users who are vulnerable, difficult to involve in the design process, or unable to partake. A specific challenge related to the application of user-centered design approaches, or co-design, emerges when users are, for various reasons, unable to participate in the design process. This challenge is at the core of this article. The proxy design method conceptualized herein can be used to involve proxy users who have either: (i) in-depth insights into actual users' backgrounds; or (ii) collective insights based on interactions with a wide variety of heterogeneous users belonging to the user group over an extended period. In our case, the roleplaying sessions initiated the co-design process when important users were unreachable. Based on our findings, derived from the empirical data regarding the design of a digital infrastructure at a cancer rehabilitation clinic, we outline the following six design guidelines to be considered when applying the proxy design method: (i) proxy design is a way of gaining an initial understanding of users to develop personas and user stories; (ii) proxy design is a way of collecting the information needed to design and develop the targeted digital artifacts; (iii) proxy design is a way of efficiently gaining an in-depth understanding of a user group that is complex, heterogeneous, has a wide range of problems, and is unreachable or unable to participate; (iv) proxy design is a way of starting the design process before involving the actual users (in addition to their proxies); (v) proxy design is a way of reducing participation overload for frail users; and (vi) proxy design is a way of facilitating collegial exchange and mutual competence development for the proxies involved in the roleplaying sessions. Lastly, we would like to stress that proxy design is a method for effectively initiating the design process when a user group is unable to participate; however, it is not a method for excluding a particular user group from the co-design effort.

To sum up, our main contribution is the conceptualization of proxy design as a method that can ignite and initiate the co-design process when important users are unreachable or unable to represent themselves in the co-design process. More specifically, we describe proxy design based on roleplaying as a fitting way of initiating the design process and consider proxy design an embedded element of co-design. Our empirical findings are based on a design ethnography that involved nurses as proxy users to speak on behalf of patients (i.e., the actual user group). We maintain that proxy design is a method that can be incorporated into co-design. Consequently, proxy design as a method allows for the involvement of users as proxies for users who are unable to represent themselves in co-design.